\documentclass[aps,showpacs,preprintnumbers,amsmath, amssymb]{revtex4}

\usepackage{float}
\usepackage{graphics,epsfig}
\usepackage{graphicx}
\usepackage{dcolumn}
\usepackage{bm}
\usepackage{color}

\begin{document}
\baselineskip=0.8 cm
\title{{\bf An analytic study on the excited states of holographic superconductors}}

\author{Xiongying Qiao$^{1}$, Dong Wang$^{1}$, Liang OuYang$^{1}$, Mengjie Wang$^{1}$\footnote{mjwang@hunnu.edu.cn}, Qiyuan Pan$^{1,2}$\footnote{panqiyuan@hunnu.edu.cn}, and Jiliang Jing$^{1,2}$\footnote{jljing@hunnu.edu.cn}}
\affiliation{$^{1}$Key Laboratory of Low Dimensional Quantum Structures and Quantum Control of Ministry of Education, Synergetic Innovation Center for Quantum Effects and Applications, and Department of Physics, Hunan Normal University, Changsha, Hunan
410081, China} \affiliation{$^{2}$Center for Gravitation and Cosmology, College of Physical Science and Technology, Yangzhou University, Yangzhou 225009, China}

\vspace*{0.2cm}
\begin{abstract}
\baselineskip=0.6 cm
\begin{center}
{\bf Abstract}
\end{center}

Based on the Sturm-Liouville eigenvalue problem, we develop a general analytic technique to investigate the excited states of the holographic superconductors. By including more higher order terms in the expansion of the trial function, we observe that the analytic results agree well with the numeric data, which indicates that the Sturm-Liouville method is very powerful to study the holographic superconductors even if we consider the excited states. For both the holographic s-wave and p-wave models, we find that the excited state has a lower critical temperature than the corresponding ground state and the difference of the dimensionless critical chemical potential between the consecutive states is around 5. Moreover, we analytically confirm that the holographic superconductor phase transition with the excited states belongs to the second order, which can be used to back up the numerical findings for both s-wave and p-wave superconductors.

\end{abstract}
\maketitle
\newpage
\vspace*{0.2cm}

\section{Introduction}

Superconductivity, which was first discovered in 1911 by Onnes \cite{Onnes}, is one of the most remarkable phenomena observed in physics in the 20th century. As a universal basis for describing the superconductivity, the Bardeen-Cooper-Schrieffer (BCS) theory \cite{BCS} may be used to explain the conventional low-$T_{c}$ superconductors by the mechanism pairing electrons but does not apply to the high-$T_{c}$ superconductors for which the strong coupling is involved. Interestingly, recent efforts showed that the anti-de Sitter/conformal field theory (AdS/CFT) correspondence \cite{Maldacena,Gubser1998,Witten}, which relates a weak coupling gravity theory in an AdS space to a strong coupling conformal field theory in one less dimension, can be used to understand the high-$T_{c}$ superconductivity \cite{HartnollRev,HerzogRev,HorowitzRev,CaiRev}. It was suggested that the spontaneous $U(1)$ symmetry breaking by bulk black holes can mimic the superconductor/conductor phase transition in the dual CFTs \cite{GubserPRD78}. The so-called holographic s-wave superconductor was first constructed by considering an Abelian Higgs model coupled to the gravity theory with a negative cosmological constant \cite{HartnollPRL101,HartnollJHEP12}, which reproduces characteristic properties shared by real superconductors. Then, the holographic p-wave superconductor was introduced in the Yang-Mills theory \cite{Gubser-Pufu} or the Maxwell complex vector field model \cite{CaiPWave-1}, and the holographic d-wave superconductor was built by introducing a charged massive spin two field propagating in the bulk \cite{DWaveChen,DWaveBenini}. It should be noted that, as suggested in \cite{CaiPWave-2} for the p-wave superconductivity, the complex vector field model is a generalization of the $SU(2)$ Yang-Mills model with a general mass and gyromagnetic ratio.

The aforementioned works on the holographic superconductors only focus on the ground state, which is the first state to condense \cite{WangSPJ}. It is of great interest to generalize the investigation to the holographic superconductors with the excited states since the excited state in superconducting materials is important in condensed matter systems \cite{PRB1988,PRB1999,RMP2004,PRL2016,LiWWZ}. In Ref. \cite{GubserPRL2008}, Gubser studied the colorful horizons with the charge in the AdS space and showed the existence of branches of solutions with multiple nodes corresponding to the excited states. More recently, Wang \emph{et al.} constructed the novel numerical solutions of the holographic s-wave superconductors with the excited states in the probe limit where the backreaction of matter fields on the spacetime metric is neglected \cite{WangJHEP2020}, and argued that the excited states of the holographic superconductors could be related to the metastable states of the mesoscopic superconductors \cite{Peeters2000,Vodolazov2002}. It was found that the excited state has a lower critical temperature $T_{c}$ than the corresponding ground state, and the conductivity $\sigma(\omega)$ of each excited state has an additional pole in Im$[\sigma(\omega)]$ and a delta function in Re$[\sigma(\omega)]$ arising at the low temperature inside the gap \cite{WangJHEP2020}. By considering the backreaction, the authors of Ref. \cite{WangLLZ} analyzed the effect of the backreaction on the condensates and optical conductivity in the excited states. In order to support numerical results and gain more physical insights for the excited states of the holographic superconductors, a fully analytic study is called for, just as pointed out in \cite{WangJHEP2020}, ``\textit{..., the difference of the critical chemical potential $\mu_{c}$ between the consecutive states is about 5 for both of the condensates $\langle O_{1}\rangle$ and $\langle O_{2}\rangle$, but the reason of these configurations is not clear. It would be very interesting to study these cases with the semi-analytical method [G. Siopsis and J. Therrien, J. High Energy Phys. {\bf 05}, 013 (2010)] to see how these values are related to excited states}".

Therefore, based on the variational method for the Sturm-Liouville eigenvalue problem first proposed in \cite{Siopsis,SiopsisBF} and later generalized to investigate holographic insulator/superconductor phase transition in \cite{CaiLZ,PanJingWang}, in this work we develop a more general analytic technique to study the excited states of the holographic s-wave superconductors in the probe limit, which will provide an explicit and complete understanding of the scalar condensate analytically. On the other hand, considering recent interests arising in the study of the new holographic p-wave superconductors via the Maxwell complex vector field model \cite{CaiPWave-1,CaiPWave-2,HuangSCPMA}, we extend the investigation to a novel family of solutions of the holographic p-wave superconductors with the excited states which, as far as we know, have not been constructed. We find that, by including more higher order terms in the expansion of the trial function, the analytic formulas obtained by the Sturm-Liouville method are in very good agreement with the numerical results both for the holographic s-wave and p-wave models with the excited states, implying that the Sturm-Liouville method is still powerful to study the holographic superconductors even if the excited states are taken into account.

This paper is organized as follows. In Sec. II we investigate the excited states of the holographic s-wave superconductors \textit{analytically}, by using the generalized Sturm-Liouville method. In particular, we calculate the critical chemical potential of the system as well as the condensate of the scalar operator near the critical point. In Sec. III, following exactly the same procedures implemented in the previous section, we discuss the p-wave models. We conclude in the last section with our main results.

\section{Excited states of the holographic s-wave superconductors}

In order to study the excited states of the holographic superconductors in the probe limit, we start with the four-dimensional planar Schwarzschild-AdS black hole
\begin{eqnarray}\label{BHMetric}
ds^2=-f(r)dt^{2}+\frac{dr^2}{f(r)}+r^{2}(dx^{2}+dy^{2})\,,
\end{eqnarray}
where $f(r)=r^2(1-r_{+}^{3}/r^{3})/L^2$. Here $L$ is the AdS radius, and $r_+$ is the radius of the event horizon. Then the Hawking temperature of the black hole is given by $T=3r_{+}/(4\pi L^2)$, which, according to the AdS/CFT correspondence, is also interpreted as the temperature of the dual system.

In the above mentioned background, we construct the holographic s-wave superconductors by considering a Maxwell field coupled with a charged complex scalar field via the action
\begin{eqnarray}\label{SWaveSystem}
S=\int d^{4}x\sqrt{-g}\left[
-\frac{1}{4}F_{\mu\nu}F^{\mu\nu}-|\nabla\psi - iA\psi|^{2}
-m^2|\psi|^2 \right] \,.
\end{eqnarray}
By taking the ansatz for the matter fields $\psi=\psi(r)$ and $A=\phi(r) dt$, we then obtain the following equations of the motion from the above action
\begin{eqnarray}
\psi^{\prime\prime}+\left(\frac{2}{r}+\frac{f^\prime}{f}\right)\psi^\prime
+\left(\frac{\phi^2}{f^2}-\frac{m^2}{f}\right)\psi=0\,, \label{SWavePsi}
\end{eqnarray}
\begin{eqnarray}
\phi^{\prime\prime}+\frac{2}{r}\phi^\prime-\frac{2\psi^2}{f}\phi=0\,, \label{SWavePhi}
\end{eqnarray}
where the prime denotes the derivative with respect to $r$. In order to solve these equations, one has to impose proper boundary conditions both at the horizon and at infinity. At the horizon $r_{+}$, we impose the boundary conditions by requiring that the scalar field $\psi$ is regular and the gauge field $A_{\mu}$ satisfies $\phi(r_{+})=0$. At infinity, the scalar and Maxwell fields behave as
\begin{eqnarray}
\psi=\frac{\psi_{-}}{r^{\Delta_{-}}}+\frac{\psi_{+}}{r^{\Delta_{+}}}\,,\hspace{0.5cm}
\phi=\mu-\frac{\rho}{r}\,, \label{infinity}
\end{eqnarray}
where $\Delta_\pm=\frac{1}{2}\left(3\pm\sqrt{9+4m^{2}L^2}\right)$ is the characteristic exponent with the Breitenlohner-Freedman (BF) bound $m^{2}_{BF}=-9/(4L^{2})$ \cite{Breitenloher}, $\mu$ and $\rho$ are interpreted as the chemical potential and charge density in the dual field theory, respectively. According to the AdS/CFT correspondence, provided $\Delta_{-}$ is larger than the unitarity bound, both $\psi_{-}$ and $\psi_{+}$ can multiply normalizable modes of the scalar field equations and correspond to the vacuum expectation values $\langle O_{-}\rangle=\sqrt{2}\psi_{-}$, $\langle O_{+}\rangle=\sqrt{2}\psi_{+}$ of an operator $O_{i}$ ($i=+$ or $-$) dual to the scalar field, respectively. We will impose boundary conditions that either $\psi_{-}$ or $\psi_{+}$ vanishes \cite{HartnollPRL101,HartnollJHEP12}.

For mathematical convenience, in the following calculations we change the variable from $r$ to $z=r_{+}/r$ with the range $0\leq z\leq1$. Then Eqs. (\ref{SWavePsi}) and (\ref{SWavePhi}) turn into
\begin{eqnarray}
\psi^{\prime\prime}+\frac{f^\prime}{f}\psi^\prime
-\frac{1}{z^{4}}\left[\frac{m^2}{f}-\frac{1}{f^2}\left(\frac{\phi}{r_{+}}\right)^{2}\right]\psi=0\,,
\label{SWavePsi-Z}
\end{eqnarray}
\begin{eqnarray}
\phi^{\prime\prime}-\frac{2\psi^2}{z^4f}\phi=0, \label{SWavePhi-Z}
\end{eqnarray}
with $f=(1/z^{2}-z)/L^{2}$. Note that here the prime denotes the derivative with respect to $z$.

\subsection{Critical chemical potential}

It has been shown numerically that, for both the ground and excited states of the holographic superconductor, there exists a critical chemical potential $\mu_{c}$ (a critical temperature $T_{c}$), above (below) which the scalar field begins to condensate due to the spontaneously broken $U(1)$ gauge symmetry \cite{WangJHEP2020}. Since the scalar field $\psi=0$ at the critical chemical potential $\mu_{c}$, as $\mu\rightarrow\mu_{c}$ below the critical point, Eq. (\ref{SWavePhi-Z}) reduces to $\phi^{\prime\prime}=0$, which has a solution
\begin{eqnarray}
\phi(z)=\mu(1-z)=\lambda r_{+c}(1-z)\,. \label{SWavePhiSolution}
\end{eqnarray}
Here we have introduced a dimensionless quantity $\lambda\equiv\mu/r_{+c}=\rho/r^{2}_{+c}$, where $r_{+c}$ is the radius of the horizon at the critical point.

According to the asymptotical behavior given by Eq. \eqref{infinity}, we take the scalar field as
\begin{eqnarray}\label{SWavePhiFz}
\psi(z)\sim \frac{\langle O_{i}\rangle}{\sqrt{2}r_{+}^{\Delta_{i}}} z^{\Delta_i}F(z)\,,
\end{eqnarray}
with the boundary condition $F(0)=1$. By inserting Eqs.~\eqref{SWavePhiSolution} and~\eqref{SWavePhiFz} into Eq. \eqref{SWavePsi-Z}, we obtain the equation of motion for the trial function $F(z)$
\begin{eqnarray}\label{SWaveFzmotion}
(TF^{\prime})^{\prime}+T\left[U+V\left(\frac{\mu}{r_{+}}\right)^{2}\right]F=0\,,
\end{eqnarray}
near the critical point, and where we have defined
\begin{eqnarray}\label{SWaveTUV}
T(z)=\frac{z^{2(\Delta_{i}-1)}(1-z^{3})}{L^{2}}\,,\;\;\;\;\;\;U(z)=\frac{\Delta_{i}}{z}\left(\frac{\Delta_{i}-1}{z}+\frac{f^\prime}{f}\right)-\frac{m^2}{z^{4}f}\,,\;\;\;\;\;\;V(z)=\frac{(1-z)^2}{z^{4}f^{2}}.
\end{eqnarray}
By making use of the Sturm-Liouville approach~\cite{Gelfand-Fomin}, the eigenvalue $\mu/r_{+}$ may be achieved from the extremal values of the following function by virtue of the Rayleigh Quotient
\begin{eqnarray}\label{SWaveEigenvalue}
\left(\frac{\mu}{r_{+}}\right)^{2}=\lambda^{2}=\frac{\int^{1}_{0}T\left(F'^{2}-UF^{2}\right)dz}{\int^{1}_{0}TVF^{2}dz}\,,
\end{eqnarray}
where we have employed the boundary condition
\begin{eqnarray}\label{TFF}
[T(z)F(z)F'(z)]|_{0}^{1}=0-T(0)F(0)F'(0)=0\,.
\end{eqnarray}
To be specific, we take $m^{2}L^{2}=-2$ for the calculations we conducted in this section. It is obvious that for this case, the condition $T(0)F(0)F'(0)=0$ is not satisfied automatically for the operator $O_{1}$ since the leading order contribution from $T(z)$ is $2(\Delta_{-}-1)=0$, as $z\rightarrow0$; while the aforementioned condition is satisfied automatically for the operator $O_{2}$ since $2(\Delta_{+}-1)=2>1$. Thus, as discussed in Refs. \cite{HFLi,WangSPJ,LvPLB2020}, we have to impose an additional Neumann boundary condition $F'(0)=0$ for the operator $O_{1}$ but need not impose any restrictions on $F'(z)$ for the operator $O_{2}$.

In order to analytically investigate the excited states of the holographic superconductors by using the Sturm-Liouville method, we shall include more higher orders of $z$ in the trial function $F(z)$. As a concrete example, we calculate the case for the operator $O_{1}$. From the above discussion, we should impose the Neumann boundary condition $F'(0)=0$ and choose the third order trial function
\begin{eqnarray}\label{SWaveTrialFunction}
F(z)=1-a_{2}z^{2}-a_{3}z^{3},
\end{eqnarray}
which leads to
\begin{eqnarray}
\begin{split}
\left(\frac{\mu}{r_{+}}\right)^{2}=60\left(\frac{1}{2}-\frac{a_{2}}{2}+\frac{5a_{2}^{2}}{6}-\frac{2a_{3}}{5}
+\frac{11a_{2}a_{3}}{7}+\frac{4a_{3}^{2}}{5}\right)/\left[-30\ln3+(60\ln3-180)a_{2}+(60\ln3-65)a_{2}^{2}\right.\nonumber\\
\left.+(40+60\ln3)a_{3}-(60\ln3-176)a_{2}a_{3}-(21+30\ln3)a_{3}^{2}
+10\sqrt{3}(1-a_{3})(1+2a_{2}-a_{3})\pi\right].
\end{split}
\end{eqnarray}
Computing the extremal values of the above expression, we can obtain the dimensionless critical chemical potential from the ground state to the second excited state, i.e., $\mu_{c}^{(0)}/r_{+}=1.121$ at $a_{2}=0.496$ and $a_{3}=-0.271$, $\mu_{c}^{(1)}/r_{+}=6.714$ at $a_{2}=15.632$ and $a_{3}=-12.681$, and $\mu_{c}^{(2)}/r_{+}=18.478$ at $a_{2}=39.772$ and $a_{3}=-57.299$. Comparing with the numerical results in \cite{WangJHEP2020}, we only obtain the first three lowest-lying modes by using the third order trial function $F(z)$.

\begin{figure}[ht]
\includegraphics[scale=0.90]{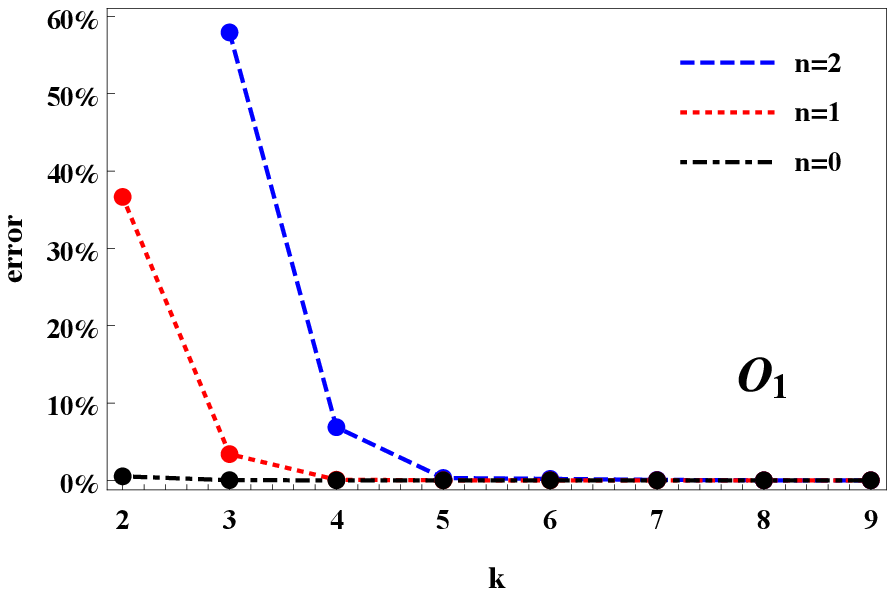}\vspace{0.0cm}
\includegraphics[scale=0.90]{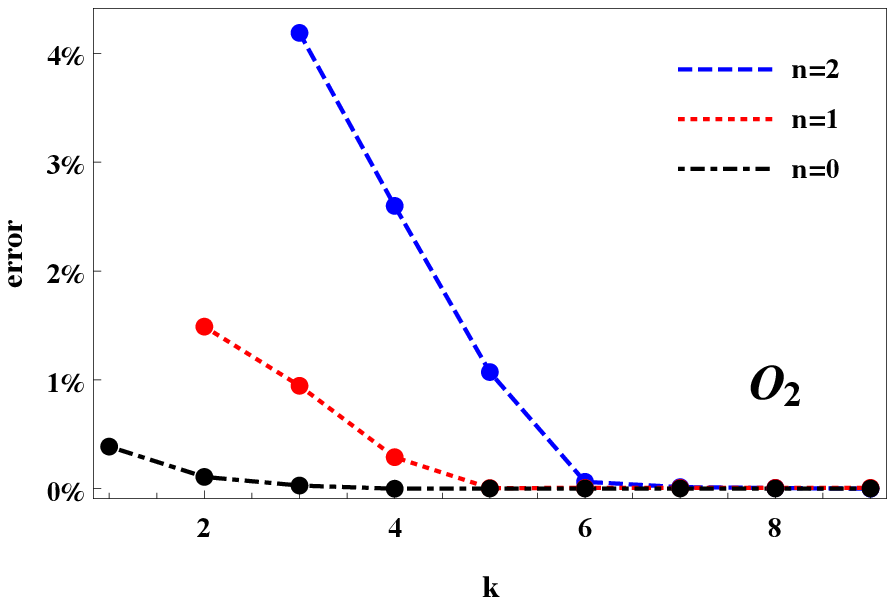} \\ \hspace{0.0cm}
\caption{\label{S-waveError} (color online) The percentage errors in the analytic estimation for the critical chemical potential of the scalar operators $O_{1}$ (left) and $O_{2}$ (right) versus the order of the expansion with the mass of the scalar field $m^2L^2=-2$. The three lines from bottom to top correspond to the ground ($n=0$, black), first ($n=1$, red) and second ($n=2$, blue) states, respectively. }
\end{figure}

\begin{figure}[ht]
\includegraphics[scale=0.90]{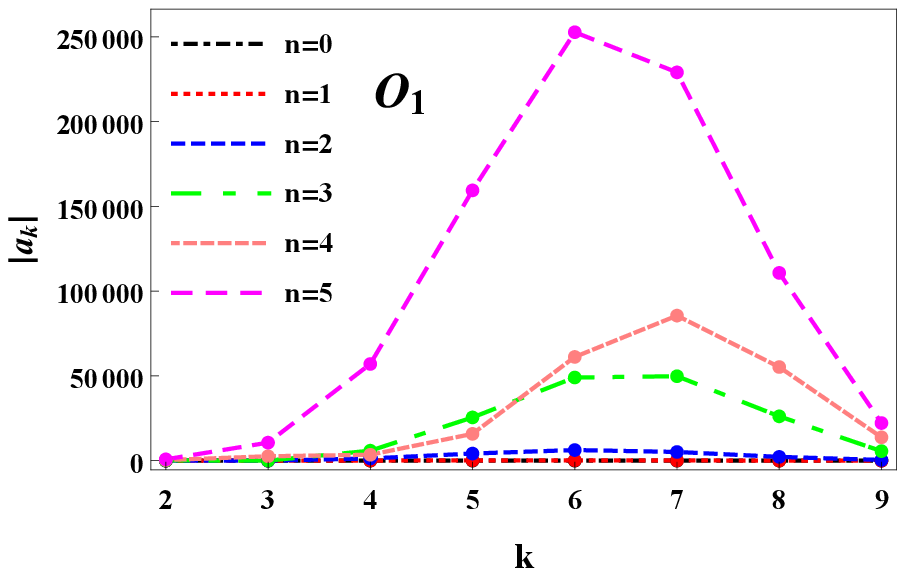}\vspace{0.0cm}
\includegraphics[scale=0.90]{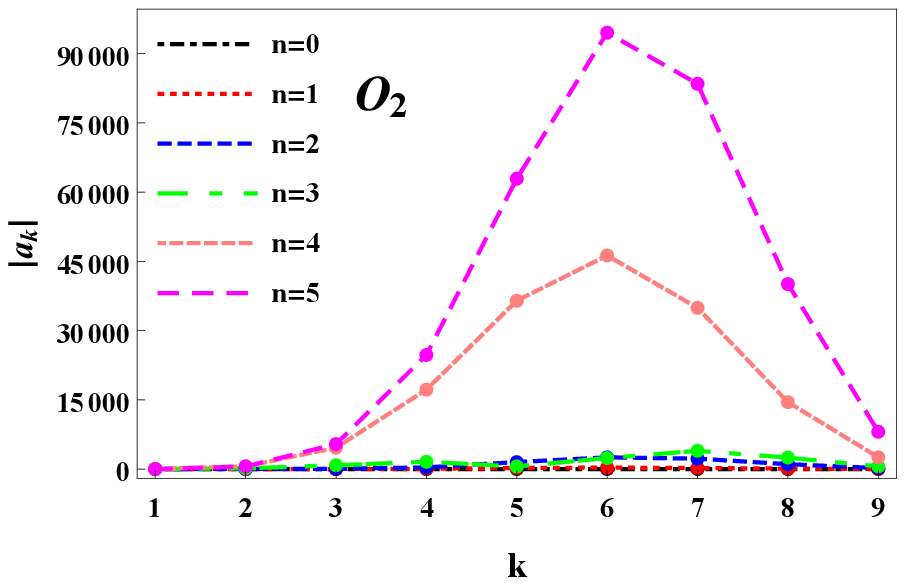} \\ \hspace{0.0cm}
\caption{\label{S-waveAk} (color online) The absolute value of $a_{k}$ as a function of the order of the expansion for the scalar operators $O_{1}$ (left) and $O_{2}$ (right) with the mass of the scalar field $m^2L^2=-2$. The six lines from bottom to top correspond to the ground ($n=0$, black), first ($n=1$, red), second ($n=2$, blue), third ($n=3$, green), fourth ($n=4$, pink) and fifth ($n=5$, magenta) states, respectively.}
\end{figure}

In order to give the higher excited state by using the analytic Sturm-Liouville method, we include the ninth order of $z$ in the trial function $F(z)$, i.e., $F(z)=1-\sum_{k=2}^{k=9}a_{k}z^{k}$ for the operator $O_{1}$, and $F(z)=1-\sum_{k=1}^{k=9}a_{k}z^{k}$ for the operator $O_{2}$. As a matter of fact, just as shown in Fig. \ref{S-waveError} for the first three lowest-lying modes, the percentage errors in the analytic estimation for the critical chemical potential of the scalar operators $O_{1}$ and $O_{2}$ drop quickly with the order of the expansion, which indicates that the trial function $F(z)$ with the ninth order of $z$ will lead to a more precise estimation. The dimensionless critical chemical potential $\mu_{c}/r_{+}$ and corresponding value of $a_{k}$ from the ground state to the fifth excited state are tabulated in Tables~\ref{SWaveO1Table} and~\ref{SWaveO2Table}. Moreover, to check the convergence of the high order expansion directly, in Fig. \ref{S-waveAk} we also exhibit the value of $|a_{k}|$ as a function of $k$ for the scalar operators $O_{1}$ and $O_{2}$, which implies that the expansion of the trial function is convergent. Comparing with the analytical result obtained from the third order trial function $F(z)$ in Eq. (\ref{SWaveTrialFunction}), we find that the value of $\mu_{c}/r_{+}$ with the ninth order trial function is much closer to the numerical result given in Tables \ref{SWaveO1Table} and \ref{SWaveO2Table}, even in the fifth excited state. The tiny error indicates that the Sturm-Liouville method with the higher order of $z$ in the trial function $F(z)$ can not only find the most stable mode, but also can find the metastable modes just like the spectral method did in \cite{WangJHEP2020}.

\begin{table}[ht]
\begin{center}
\caption{\label{SWaveO1Table}
The dimensionless critical chemical potential $\mu_{c}/r_{+}$ for the operator $O_{1}$ and corresponding value of $a_{k}$ for the trial function $F(z)=1-\sum_{k=2}^{k=9}a_{k}z^{k}$ in the holographic s-wave superconductor. The results of $\mu_{c}/r_{+}$ are obtained analytically by the Sturm-Liouville method (left column) and numerically by the spectral method \cite{WangJHEP2020} (right column) from the ground state to the fifth excited state.}
\begin{tabular}{c |c| c c c c c c c c c}
\hline
 $n$~&~$\mu_{c}/r_{+}$~&~$a_{2}$~&~$a_{3}$~&~$a_{4}$~&~$a_{5}$~&~$a_{6}$~&~$a_{7}$~&~$a_{8}$~&~$a_{9}$  \\
\hline
$0$ &~1.120~~1.120~&~0.628~&~-0.584~&~0.031~&~0.507~&~-0.636~&~0.412~&~-0.148~&~0.023 \\
\hline
$1$ &~6.493~~6.494~&~21.006~&~-12.433~&~-87.517~&~217.020~&~-242.872~&~150.734~&~-49.962~&~6.827 \\
\hline
$2$ &~11.700~~11.701~&~66.377~&~7.206~&~-1310.093~&~4208.241~&~-6279.568~&~5139.829~&~-2245.136~&~411.703 \\
\hline
$3$ &~16.901~~16.898~&~142.425~&~29.025~&~-5888.970~&~25509.602~&~-49071.741~&~49823.574~&~-26164.029~&~5624.096 \\
\hline
$4$ &~22.258~~22.094~&~370.979~&~-2616.084~&~3486.249~&~15827.106~&~-61194.858~&~85570.988~&~-55240.956~&~13794.571 \\
\hline
$5$ &~28.055~~27.290~&~804.021~&~-10643.884~&~57037.950~&~-159446.972~&~252736.492~&~-229055.289~&~110764.290~&~-22192.871 \\
\hline
\end{tabular}
\end{center}
\end{table}

\begin{table}[ht]
\begin{center}
\caption{\label{SWaveO2Table}
The dimensionless critical chemical potential $\mu_{c}/r_{+}$ for the operator $O_{2}$ and corresponding value of $a_{k}$ for the trial function $F(z)=1-\sum_{k=1}^{k=9}a_{k}z^{k}$ in the holographic s-wave superconductor. The results of $\mu_{c}/r_{+}$ are obtained analytically by the Sturm-Liouville method (left column) and numerically by the spectral method \cite{WangJHEP2020} (right column) from the ground state to the fifth excited state.}
\begin{tabular}{c |c| c c c c c c c c c c}
\hline
 $n$~&~$\mu_{c}/r_{+}$~&~$a_{1}$~&~$a_{2}$~&~$a_{3}$~&~$a_{4}$~&~$a_{5}$~&~$a_{6}$~&~$a_{7}$~&~$a_{8}$~&~$a_{9}$  \\
\hline
$0$ &~4.064~~4.064~&~-0.000(1)~&~2.752~&~-3.070~&~-1.656~&~8.040~&~-10.214~&~7.040~&~-2.670~&~0.439 \\
\hline
$1$ &~9.188~~9.188~&~-0.001~&~14.001~&~-12.566~&~-73.632~&~230.420~&~-314.577~&~238.755~&~-98.493~&~17.327 \\
\hline
$2$ &~14.357~~14.357~&~-0.336~&~39.647~&~-62.264~&~-356.566~&~1509.792~&~-2537.698~&~2259.881~&~-1057.261~&~205.617 \\
\hline
$3$ &~19.546~~19.538~&~-5.002~&~161.141~&~-819.342~&~1573.056~&~-570.022~&~-2389.964~&~3955.336~&~-2487.386~&~583.362 \\
\hline
$4$ &~24.784~~24.725~&~-22.302~&~606.951~&~-4619.875~&~17210.123~&~-36466.961~&~46282.321~&~-34939.465~&~14505.174~&~-2555.152 \\
\hline
$5$ &~30.367~~29.915~&~-11.556~&~556.828~&~-5415.397~&~24714.429~&~-62911.944~&~94548.945~&~-83477.422~&~40083.735~&~-8086.476 \\
\hline
\end{tabular}
\end{center}
\end{table}

From Tables~\ref{SWaveO1Table} and~\ref{SWaveO2Table}, we observe that the critical chemical potential $\mu_{c}$ increases with increasing the number of nodes $n$ for both the operators $O_{1}$ and $O_{2}$, which agrees well with the fact that the ground state first appears with the decrease of the temperature and the solutions of the first-excited state begin to develop by further decreasing the temperature to the critical temperature of the first-excited state. This can be used to back up the numerical finding given in Ref. \cite{WangJHEP2020} that there exists a lower critical temperature in the corresponding excited state. Using the analytic results obtained by the Sturm-Liouville method, we can express the relation between $\mu_{c}/r_{+}$ and $n$ as
\begin{eqnarray}\label{SWaveMuc}
\frac{\mu_{c}}{r_{+}}\approx
\left\{
\begin{array}{rl}
5.347n+1.053   \ , &  \quad {\rm for} \ O_{1}\,,\\ \\
5.322n+3.840   \ , &  \quad {\rm for} \  O_{2}\,,
\end{array}\right.
\end{eqnarray}
which is in very good agreement with the numeric results given in~\cite{WangJHEP2020}, and shows that for both operators the difference of the dimensionless critical chemical potential $\mu_{c}/r_{+}$ between the consecutive states is around 5.

\subsection{Critical phenomena}

In the vicinity of the critical point, the condensate for the scalar operator $O_{i}$ is small. Therefore, we can expand $\phi(z)$ in small $\langle O_{i}\rangle$ as
\begin{eqnarray}
\frac{\phi(z)}{r_{+}}=\lambda(1-z)+\frac{\langle O_{i}\rangle^{2}}{r_{+}^{2\Delta_{i}}}\chi(z)+\cdots,
\end{eqnarray}
with the boundary condition $\chi(1)=\chi'(1)=0$ at the event horizon \cite{Siopsis}. Substituting the above expression and function (\ref{SWavePhiFz}) into Eq. (\ref{SWavePhi-Z}), we obtain the equation of motion for $\chi(z)$
\begin{eqnarray}\label{SWaveChi}
\chi''-\lambda\frac{z^{2(\Delta_{i}-2)}(1-z)F^2}{f}=0.
\end{eqnarray}
Making integration for both sides of the above equation, we obtain
\begin{eqnarray}\label{SWaveChiPrime0}
\chi'(0)=-\lambda C_{i}=-\lambda\int^{1}_{0}\frac{z^{2(\Delta_{i}-2)}(1-z)F^2}{f}dz.
\end{eqnarray}

According to the asymptotic behavior given in Eq. (\ref{infinity}), we can expand $\phi(z)$ near $z\rightarrow0$ as
\begin{eqnarray}\label{ExpandingSWavePhi}
\frac{\phi(z)}{r_{+}}=\frac{\rho}{r_{+}^{2}}(1-z)=\lambda(1-z)+\frac{\langle O_{i}\rangle^{2}}{r_{+}^{2\Delta_{i}}}\left[\chi(0)+\chi'(0)z+\cdots\right].
\end{eqnarray}
From the coefficients of the $z^{1}$ term, we have
\begin{eqnarray}\label{SWaveCondensate}
\frac{\langle O_{i}\rangle}{T_{c}^{\Delta_{i}}}\approx\sqrt{\frac{2}{C_{i}}}\left(\frac{4\pi}{3}\right)^{\Delta_{i}}
\left(1-\frac{T}{T_{c}}\right)^{1/2},
\end{eqnarray}
where the critical temperature is given by
\begin{eqnarray}\label{SWaveTc}
T_{c}=\frac{3}{4\pi}\left(\frac{\rho}{\lambda_{ext}}\right)^{1/2},
\end{eqnarray}
with the extremal values $\lambda_{ext}$ of the expression (\ref{SWaveEigenvalue}). From Eq.~\eqref{SWaveCondensate}, it is shown clearly that the phase transition of the holographic s-wave superconductors belongs to the second order and the critical exponent of the system takes the mean field value $1/2$, even in the excited states.

\begin{figure}[ht]
\includegraphics[scale=0.90]{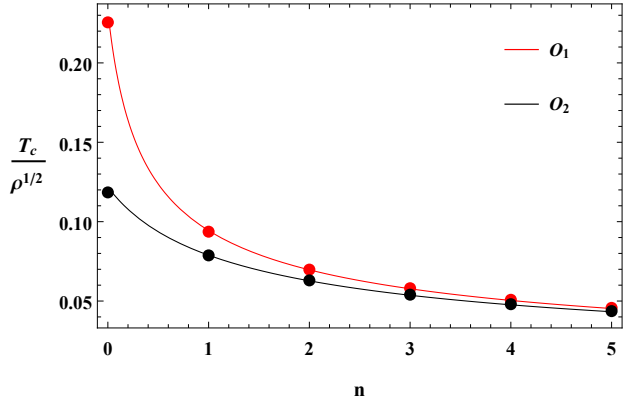} \hspace{0.0cm}
\caption{\label{S-waveTc} (color online) The dimensionless critical temperature $T_{c}/\rho^{1/2}$ as a function of the number
of nodes $n$ for the scalar operators $O_{1}$ (red) and $O_{2}$ (black) with the mass of the scalar field $m^2L^2=-2$. The data points represent the numerical results and the solid lines are obtained by using the analytic expression (\ref{SWaveMuc}).}
\end{figure}

To be specific, we take the scalar field mass as $m^{2}L^{2}=-2$ in the holographic s-wave superconductors.
We have calculated the first three lowest-lying modes, for the operator $O_{1}$
\begin{eqnarray}\label{SWaveO1}
\langle O_{1}\rangle\approx
\left\{
\begin{array}{rl}
8.2T_{c}^{(0)}(1-T/T_{c}^{(0)})^{1/2}   \ , &  \quad {\rm the~ground~state~with}  \  T_{c}^{(0)}\approx0.226\rho^{1/2}\,, \\ \\
6.6T_{c}^{(1)}(1-T/T_{c}^{(1)})^{1/2}   \ , &  \quad {\rm the~1st~excited~state~with}  \  T_{c}^{(1)}\approx0.094\rho^{1/2}\,, \\ \\ 6.2T_{c}^{(2)}(1-T/T_{c}^{(2)})^{1/2}   \ , &  \quad {\rm the~2nd~excited~state~with}  \  T_{c}^{(2)}\approx0.070\rho^{1/2}\,,
\end{array}\right.
\end{eqnarray}
and for the operator $O_{2}$
\begin{eqnarray}\label{SWaveO2}
\langle O_{2}\rangle\approx
\left\{
\begin{array}{rl}
119(T_{c}^{(0)})^{2}(1-T/T_{c}^{(0)})^{1/2}   \ , &  \quad {\rm the~ground~state~with}  \  T_{c}^{(0)}\approx0.118\rho^{1/2}\,, \\ \\
245(T_{c}^{(1)})^{2}(1-T/T_{c}^{(1)})^{1/2}   \ , &  \quad {\rm the~1st~excited~state~with}  \  T_{c}^{(1)}\approx0.079\rho^{1/2}\,, \\ \\ 364(T_{c}^{(2)})^{2}(1-T/T_{c}^{(2)})^{1/2}   \ , &  \quad {\rm the~2nd~excited~state~with}  \  T_{c}^{(2)}\approx0.063\rho^{1/2}\,,
\end{array}\right.
\end{eqnarray}
and both of which agree well with the numerical data obtained by the spectral method in \cite{WangJHEP2020}, especially the critical temperatures $T_{c}$. For the ground state, comparing with the analytical results from the second order trial function $F(z)$ \cite{Siopsis,ZengGao,GangopadhyayR,PanJWC}, i.e., $T_{c}^{(0)}\approx0.225\rho^{1/2}$ for the operator $O_{1}$ and $T_{c}^{(0)}\approx0.117\rho^{1/2}$ for the operator $O_{2}$, we observe that the values of $T_{c}$ with the ninth order trial function are much more closer to the numerical findings. On the other hand, using the analytic expression (\ref{SWaveMuc}) and numerical results, we plot the values of the dimensionless critical temperature $T_{c}/\rho^{1/2}$ in function of $n$ for the scalar operators $O_{1}$ and $O_{2}$ in Fig. \ref{S-waveTc}, which shows that the critical temperature decreases with the increase of $n$. This means that the excited state has a lower critical temperature than the corresponding ground state.

Through the above analytic calculations we have confirmed, on one hand, that the condensate of the excited states is smaller (larger) than that of the ground state for the operator $O_{1}$ ($O_{2}$); on the other hand, that the Sturm-Liouville method is a very powerful approach to analytically investigate the holographic s-wave superconductors for both the ground and excited states.

\section{Excited states of the holographic p-wave superconductors}

Since the Sturm-Liouville method is effective to analyze the excited states of the holographic s-wave superconductors, in this section, we extend it to analytically study the excited states of the holographic p-wave superconductors which, as far as we know, has not been explored yet.

In the probe limit, the holographic p-wave superconductors can be constructed via the Maxwell complex vector field model \cite{CaiPWave-1,CaiPWave-2}
\begin{eqnarray}\label{NewPWction}
S=\int
d^{4}x\sqrt{-g}\left(-\frac{1}{4}F_{\mu\nu}F^{\mu\nu}-\frac{1}{2}\rho_{\mu\nu}^{\dag}\rho^{\mu\nu}-m^2\rho_{\mu}^{\dag}\rho^{\mu}+i
q\gamma\rho_{\mu}\rho_{\nu}^{\dag}F^{\mu\nu}\right)\,,
\end{eqnarray}
where the tensor $\rho_{\mu\nu}$ is defined by
$\rho_{\mu\nu}=D_\mu\rho_\nu-D_\nu\rho_\mu$ with $D_\mu=\nabla_\mu-iqA_\mu$ being the covariant derivative, $m$ and $q$ are the mass and charge of the vector field $\rho_\mu$, respectively. It should be noted that, the last term $i
q\gamma\rho_{\mu}\rho_{\nu}^{\dag}F^{\mu\nu}$, which describes the interaction between the vector field $\rho_\mu$ and gauge field $A_\mu$, will not play a part in the present study since we will consider the case without external magnetic field. Without loss of generality, we take the unit with the field charge $q=1$ in the following calculation, following Ref. \cite{CaiPWave-1}.

By taking the ansatz for the matter fields
\begin{eqnarray}\label{PWaveAnsatz}
\rho_{\mu}dx^{\mu}=\rho_{x}(r)dx\,,~~A_\mu dx^{\mu}=A_t(r)dt\,,
\end{eqnarray}
where $\rho_{x}(r)$ and $A_t(r)$ are real functions, one may derive the
following equations of motion in the background of the Schwarzschild-AdS black hole
\begin{eqnarray}
\rho_{x}^{\prime\prime}+\frac{f^\prime}{f}\rho_{x}^{\prime}
+\left(\frac{A_{t}^2}{f^2}-\frac{m^2}{f}\right)\rho_{x}=0\,,
\label{rhoxPWave}
\end{eqnarray}
\begin{eqnarray}
A_t^{\prime\prime}+\frac{2}{r}A_t^{\prime}-\frac{2\rho_{x}^2}{r^2f}A_t=0\,, \label{AtPWave}
\end{eqnarray}
where the prime denotes the derivative with respect to $r$. Obviously, if we set $m^{2}L^{2}=0$, $A_t=\tilde{\Phi}$ and rescale the field by $\rho_{x}=\tilde{w}/\sqrt{2}$ in the above equations, we can recover the equations of motion (3.4) in \cite{Gubser-Pufu} for the holographic p-wave superconductors where an $SU(2)$ Yang-Mills action is considered. For the boundary conditions at the horizon $r_{+}$, the vector field $\rho_\mu$ is required to be regular and the gauge field $A_\mu$ obeys $A_{t}(r_{+})=0$. At infinity $r\rightarrow\infty$, the solutions of Eqs.~\eqref{rhoxPWave} and~\eqref{AtPWave} behave as
\begin{eqnarray}\label{PWInfinityCondition}
\rho_{x}=\frac{\rho_{x-}}{r^{\Delta_{-}}}+\frac{\rho_{x+}}{r^{\Delta_{+}}}\,,\hspace{0.5cm}
A_t=\mu-\frac{\rho}{r}\,,
\end{eqnarray}
where $\Delta_{\pm}=\tfrac{1}{2}(1\pm\sqrt{1+4m^{2}L^{2}})$ is again the characteristic exponent with the mass beyond the BF bound $m_{BF}^2=-1/(4L^{2})$ \cite{Breitenloher}, $\rho_{x-}$ and $\rho_{x+}$ are interpreted as the source and the vacuum expectation value of the vector operator $O_{x}$ in the dual field theory, respectively. Since we are interested in the case where the condensate appears spontaneously, we will impose boundary condition $\rho_{x-}=0$. In the following discussions, we will use $\Delta$ to denote $\Delta_{+}$ for simplicity.

Following the same procedures we employed in the previous section, in order to solve Eqs.~\eqref{rhoxPWave} and~\eqref{AtPWave} analytically, we define the variable $z=r_{+}/r$ and then Eqs.~\eqref{rhoxPWave} and~\eqref{AtPWave} turn into
\begin{eqnarray}
\rho_{x}^{\prime\prime}+\left(\frac{2}{z}+\frac{f^\prime}{f}\right)\rho_{x}^\prime
-\frac{1}{z^{4}}\left[\frac{m^2}{f}-\frac{1}{f^2}\left(\frac{A_t}{r_{+}}\right)^{2}\right]\rho_{x}=0\,, \label{rhoxPWaveZ}
\end{eqnarray}
\begin{eqnarray}
A_{t}^{\prime\prime}-\frac{2}{z^2f}\left(\frac{\rho_{x}}{r_{+}}\right)^{2}A_{t}=0\,, \label{pwave-PhiZ}
\end{eqnarray}
where the prime now denotes the derivative with respect to $z$.

\subsection{Critical chemical potential}

Similarly to the analysis provided in the previous section, if $\mu\leq\mu_{c}$, the vector field $\rho_{x}=0$. Therefore, below the critical point, Eq. (\ref{pwave-PhiZ}) becomes $A_{t}^{\prime\prime}=0$, which leads to the same solution as in (\ref{SWavePhiSolution}) for $\phi$, i.e.,
\begin{eqnarray}
A_{t}(z)=\mu(1-z)=\lambda r_{+c}(1-z)\,, \label{PWavePhiSolution}
\end{eqnarray}
with a dimensionless quantity $\lambda\equiv\mu/r_{+c}=\rho/r^{2}_{+c}$.

Taking into account the asymptotic behavior from Eq. (\ref{PWInfinityCondition}), we assume that the vector field $\rho_{x}$ takes the form
\begin{eqnarray}\label{PWavePhiFz}
\rho_{x}(z)\sim \frac{\langle O_{x}\rangle}{r_{+}^{\Delta}} z^{\Delta}F(z)\,,
\end{eqnarray}
where the trial function $F(z)$ with the boundary conditions $F(0)=1$ satisfies the following equation of motion
\begin{eqnarray}\label{PWaveFzmotion}
(PF^{\prime})^{\prime}+P\left[Q+V\left(\frac{\mu}{r_{+}}\right)^{2}\right]F=0\,,
\end{eqnarray}
with
\begin{eqnarray}\label{PWavePQV}
P(z)=\frac{z^{2\Delta}(1-z^{3})}{L^{2}}\,,\;\;\;\;\;\;Q(z)=\frac{\Delta}{z}\left(\frac{1+\Delta}{z}+\frac{f^\prime}{f}\right)-\frac{m^2}{z^{4}f}\,.
\end{eqnarray}
Here, the function $V(z)$ is defined in Eq.~\eqref{SWaveTUV}. Following the Sturm-Liouville method~\cite{Gelfand-Fomin}, the eigenvalues of $\mu/r_{+}$ can be obtained from variation of the following function
\begin{eqnarray}\label{PWaveEigenvalue}
\left(\frac{\mu}{r_{+}}\right)^{2}=\lambda^{2}=\frac{\int^{1}_{0}P\left(F'^{2}-QF^{2}\right)dz}{\int^{1}_{0}PVF^{2}dz}.
\end{eqnarray}
It should be noted that $[P(z)F(z)F'(z)]|_{0}^{1}=0$, because of the fact that $P(1)\equiv0$ and $P(0)\equiv0$ for the case of $\Delta=(1+\sqrt{1+4m^{2}L^{2}})/2$. Thus, similarly to the operator $O_{2}$ in the holographic s-wave superconductors, we require $F(z)$ to obey the Dirichlet boundary condition $F(0)=1$ rather than the Neumann boundary condition $F'(0)=0$.

\begin{figure}[ht]
\includegraphics[scale=0.9]{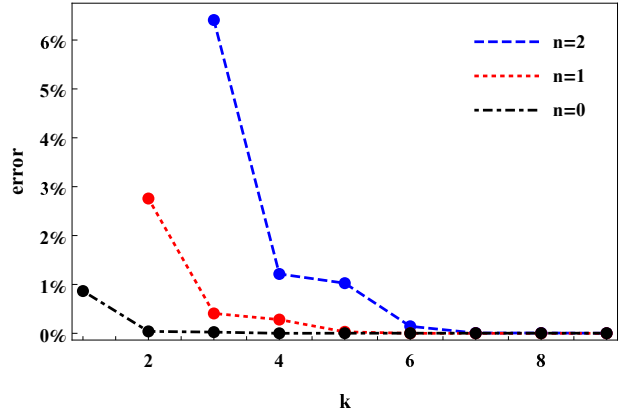}\\ \vspace{0.0cm}
\caption{\label{P-waveError} (color online) The percentage errors in the analytic estimation for the critical chemical potential of the vector operator $O_{x}$ versus the order of the expansion with the mass of the vector field $m^2L^2=3/4$. The three lines from bottom to top correspond to the ground ($n=0$, black), first ($n=1$, red) and second ($n=2$, blue) states, respectively. }
\end{figure}

\begin{figure}[ht]
\includegraphics[scale=0.90]{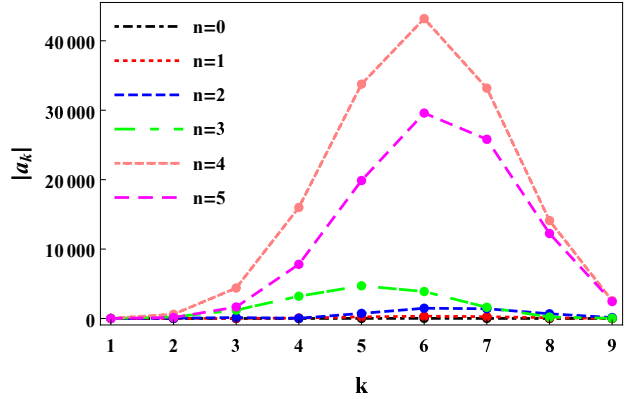} \hspace{0.0cm}
\caption{\label{P-waveAk} (color online) The absolute value of $a_{k}$ as a function of the order of the expansion for the vector operator $O_{x}$ with the mass of the vector field $m^2L^2=3/4$. The six lines correspond to the ground ($n=0$, black), first ($n=1$, red), second ($n=2$, blue), third ($n=3$, green), fourth ($n=4$, pink) and fifth ($n=5$, magenta) states, respectively.}
\end{figure}

To be specific, in the following calculations we take the mass of the vector field as $m^{2}L^{2}=3/4$, and the other choices of the mass will not change our results qualitatively. In order to obtain the higher excited states of the holographic p-wave superconductors, we include the ninth order of $z$ in the trial function $F(z)$, i.e., $F(z)=1-\sum_{k=1}^{k=9}a_{k}z^{k}$ for the operator $O_{x}$, which will lead to a more precise estimation for the critical chemical potential, just as shown in Fig. \ref{P-waveError} for the first three lowest-lying modes, where the percentage errors in the analytic estimation drop quickly with the order of the expansion. We give the dimensionless critical chemical potential $\mu_{c}/r_{+}$ and corresponding value of $a_{k}$ from the ground state to the fifth excited state in Table \ref{PWaveTable}, and plot the value of $|a_{k}|$ in function of $k$ in Fig. \ref{P-waveAk} which indicates the convergence of the high order expansion for the trial function $F(z)$. It is shown clearly that, the agreement between the analytic results derived from the Sturm-Liouville method and the numeric data obtained by the shooting method is impressive, which implies that the Sturm-Liouville method is a powerful approach to analytically investigate the excited states, not only for the s-wave but also for the p-wave, holographic superconductors.

\begin{table}[ht]
\begin{center}
\caption{\label{PWaveTable}
The dimensionless critical chemical potential $\mu_{c}/r_{+}$ for the operator $O_{x}$ and corresponding value of $a_{k}$ for the trial function $F(z)=1-\sum_{k=1}^{k=9}a_{k}z^{k}$ in the holographic p-wave superconductor. The results of $\mu_{c}/r_{+}$ are obtained analytically by the Sturm-Liouville method (left column) and numerically by the shooting method (right column) from the ground state to the fifth excited state.}
\begin{tabular}{c |c| c c c c c c c c c c}
\hline
 $n$~&~$\mu_{c}/r_{+}$~&~$a_{1}$~&~$a_{2}$~&~$a_{3}$~&~$a_{4}$~&~$a_{5}$~&~$a_{6}$~&~$a_{7}$~&~$a_{8}$~&~$a_{9}$  \\
\hline
$0$ &~5.465~~5.465~&~-0.001~&~3.748~&~-4.383~&~-3.590~&~15.443~&~-20.024~&~14.098~&~-5.429~&~0.901 \\
\hline
$1$ &~10.569~~10.569~&~-0.048~&~14.537~&~-17.447~&~-65.639~&~232.679~&~-333.649~&~260.828~&~-109.675~&~19.531 \\
\hline
$2$ &~15.723~~15.723~&~-1.113~&~47.342~&~-134.653~&~-29.151~&~735.009~&~-1478.787~&~1407.648~&~-679.655~&~134.282 \\
\hline
$3$ &~20.901~~20.894~&~-9.712~&~218.109~&~-1207.695~&~3210.790~&~-4714.401~&~3881.296~&~-1617.265~&~201.046~&~38.902 \\
\hline
$4$ &~26.104~~26.074~&~-26.182~&~609.604~&~-4393.739~&~15997.401~&~-33753.529~&~43196.599~&~-33177.421~&~14105.298~&~-2557.100 \\
\hline
$5$ &~31.984~~31.258~&~5.275~&~137.855~&~-1653.369~&~7806.518~&~-19871.633~&~29586.013~&~-25813.604~&~12246.064~&~-2442.091 \\
\hline
\end{tabular}
\end{center}
\end{table}

From Table \ref{PWaveTable}, we find that the critical chemical potential $\mu_{c}$ increases as the number of nodes $n$ increases for the operator $O_{x}$, which shows that the excited state has a lower critical temperature than the ground state, just as observed for the holographic s-wave superconductors. Fitting the relation between $\mu_{c}/r_{+}$ and $n$ by using the analytic results, we get
\begin{eqnarray}\label{PWaveMuc}
\frac{\mu_{c}}{r_{+}}\approx 5.268n+5.288,
\end{eqnarray}
which agrees well with the numeric result $\mu_{c}/r_{+}\approx 5.161n+5.427$. This behavior is reminiscent of that observed for the holographic s-wave case in Eq. (\ref{SWaveMuc}), so we conclude that the difference of the dimensionless critical chemical potential $\mu_{c}/r_{+}$ between the consecutive states is around 5 for both s-wave and p-wave holographic superconductors.

\subsection{Critical phenomena}

Considering that the condensate value of the vector field $\rho_{x}(z)$ is so small, we expand $A_{t}(z)$ in small $\langle O_{x}\rangle$ as
\begin{eqnarray}
\frac{A_{t}(z)}{r_{+}}=\lambda(1-z)+\frac{2\langle O_{x}\rangle^{2}}{r_{+}^{2(1+\Delta)}}\chi(z)+\cdots\,,
\end{eqnarray}
with the boundary condition $\chi(1)=\chi'(1)=0$. With the help of Eqs. (\ref{pwave-PhiZ}) and (\ref{PWavePhiFz}), we have
\begin{eqnarray}\label{pwavechi}
\chi''-\lambda\frac{z^{2(\Delta-1)}(1-z)F^2}{f}=0\,,
\end{eqnarray}
which leads to the expression
\begin{eqnarray}\label{PWaveChiPrime0}
\chi'(0)=-\lambda C_{x}=-\lambda\int^{1}_{0}\frac{z^{2(\Delta-1)}(1-z)F^2}{f}dz\,.
\end{eqnarray}

\begin{figure}[ht]
\includegraphics[scale=0.7]{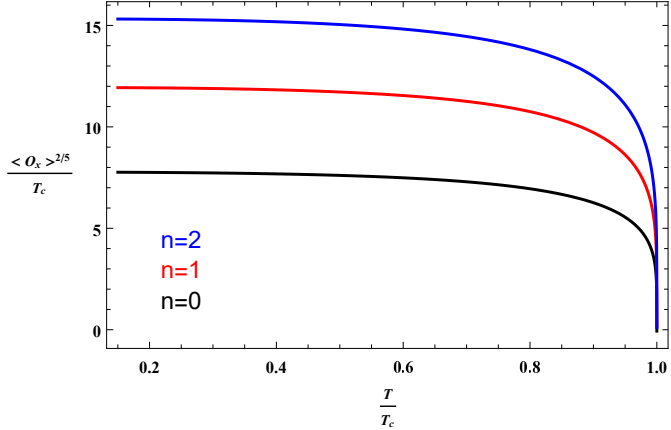}\\ \vspace{0.0cm}
\caption{\label{CondPWave} (color online) The condensates of the vector operator $O_{x}$ as a function of temperature for the mass of the vector field $m^2L^2=3/4$. The three lines from bottom to top correspond to the ground ($n=0$, black), first ($n=1$, red) and second ($n=2$, blue) states, respectively. }
\end{figure}

Near $z\rightarrow0$, $A_{t}(z)$ may be expanded as
\begin{eqnarray}\label{ExpandingPWavePhi}
\frac{A_{t}(z)}{r_{+}}=\frac{\rho}{r_{+}^{2}}(1-z)=\lambda(1-z)+\frac{2\langle O_{x}\rangle^{2}}{r_{+}^{2(1+\Delta)}}\left[\chi(0)+\chi'(0)z+\cdots\right]\,.
\end{eqnarray}
Comparing the coefficients of the $z^{1}$ term in both sides of the above formula, we obtain
\begin{eqnarray}\label{PWaveCondensate}
\frac{\langle O_{x}\rangle}{T_{c}^{1+\Delta}}\approx\frac{1}{\sqrt{C_{x}}}\left(\frac{4\pi}{3}\right)^{1+\Delta}
\left(1-\frac{T}{T_{c}}\right)^{1/2}\,,
\end{eqnarray}
where the critical temperature is given in Eq. (\ref{SWaveTc}) with the extremal values $\lambda_{ext}$ of the expression (\ref{PWaveEigenvalue}). It is obvious that the phase transition of the holographic p-wave superconductors is always of the second order and the condensate approaches zero as $\langle O_{x}\rangle\sim(T_{c}-T)^{\beta}$ with the mean field critical exponent $\beta=1/2$ for all excited states, which is consistent with the numerical finding shown in Fig. \ref{CondPWave} for the first three lowest-lying modes with $m^{2}L^{2}=3/4$.

\begin{figure}[ht]
\includegraphics[scale=0.66]{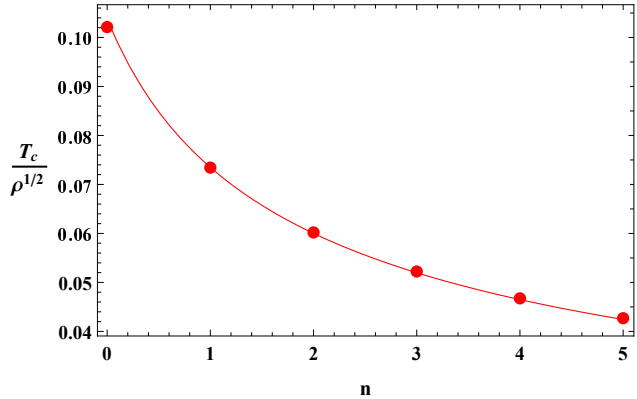} \hspace{0.0cm}
\caption{\label{P-waveTc} (color online) The dimensionless critical temperature $T_{c}/\rho^{1/2}$ as a function of the number
of nodes $n$ for the vector operator $O_{x}$ with the mass of the vector field $m^2L^2=3/4$. The data points represent the numerical results and the solid line is obtained by using the analytic expression (\ref{PWaveMuc}).}
\end{figure}

For the mass of the vector field $m^2L^2=3/4$ considered here, as $T\rightarrow T_{c}$, we obtain the first three lowest-lying modes
\begin{eqnarray}\label{PWaveOx}
\langle O_{x}\rangle\approx
\left\{
\begin{array}{rl}
277(T_{c}^{(0)})^{5/2}(1-T/T_{c}^{(0)})^{1/2}\, , & \quad {\rm the~ground~state~with}\ T_{c}^{(0)}\approx0.102\rho^{1/2}\,, \\ \\
760(T_{c}^{(1)})^{5/2}(1-T/T_{c}^{(1)})^{1/2}\, , & \quad {\rm the~1st~excited~state~with}\ T_{c}^{(1)}\approx0.073\rho^{1/2}\,, \\ \\ 1332(T_{c}^{(2)})^{5/2}(1-T/T_{c}^{(2)})^{1/2}\, , & \quad {\rm the~2nd~excited~state~with}\ T_{c}^{(2)}\approx0.060\rho^{1/2}\,,
\end{array}\right.
\end{eqnarray}
which can be compared with the numerical fitting results $\langle O_{x}^{(0)}\rangle\approx339(T_{c}^{(0)})^{5/2}(1-T/T_{c}^{(0)})^{1/2}$ with $ T_{c}^{(0)}\approx0.102\rho^{1/2}$, $\langle O_{x}^{(1)}\rangle\approx1026(T_{c}^{(1)})^{5/2}(1-T/T_{c}^{(1)})^{1/2}$ with $ T_{c}^{(1)}\approx0.073\rho^{1/2}$, and $\langle O_{x}^{(2)}\rangle\approx1926(T_{c}^{(2)})^{5/2}(1-T/T_{c}^{(2)})^{1/2}$ with $ T_{c}^{(2)}\approx0.060\rho^{1/2}$. In Fig. \ref{P-waveTc}, we present the dimensionless critical temperature $T_{c}/\rho^{1/2}$ as a function of $n$ for the vector operator $O_{x}$, which tells us that the critical temperature decreases as the number of nodes $n$ increases. On the other hand, our result shows that, similar to the behavior of the scalar operator $O_{2}$ in the holographic s-wave superconductors, the condensate of the excited state is larger than that of the ground state for the vector operator $O_{x}$. Obviously, the Sturm-Liouville method is still powerful to disclose the property of the holographic p-wave superconductors with the excited states.

\section{conclusions}

Based on the variational method for the Sturm-Liouville eigenvalue problem, we proposed a general analytic technique, by including more higher order terms in the expansion of the trial function, to investigate the excited states of the holographic superconductors in the probe limit, which provides an explicit and complete understanding of the phase transition in the holographic systems. We found that our analytic results are in very good agreement with the numeric calculation, both for the s-wave (the scalar field) model with the excited states first introduced in \cite{WangJHEP2020} and for the p-wave (the vector field) one newly constructed in this work, which implies that the Sturm-Liouville method is a powerful approach to study the holographic superconductors, even when the excited states are taken into account. It was observed that, both for the holographic s-wave and p-wave superconductors, the dimensionless critical chemical potential $\mu_{c}/r_{+}$ increases with increasing the number of nodes, indicating that the excited state has a lower critical temperature than the corresponding ground state. Interestingly, although the underlying mechanism remains mysterious, we observed that the difference of the dimensionless critical chemical potential between the consecutive states is around 5 regardless of the s-wave or p-wave superconductors. Moreover, with the help of the analytic Sturm-Liouville method, we presented the condensates of the excited states near the critical point and found that the number of nodes \textit{does not} modify the critical phenomena, i.e., the holographic superconductor phase transition belongs to the second order and the critical exponent of the system always takes the mean-field value, which can be used to back up the numerical findings for both s-wave and p-wave superconductors with the excited states.

\begin{acknowledgments}

This work was supported by the National Natural Science Foundation of China under Grant Nos. 11775076, 11875025, 11705054 and 11690034; Hunan Provincial Natural Science Foundation of China under Grant Nos. 2018JJ3326 and 2016JJ1012.

\end{acknowledgments}

\end{document}